\documentstyle[12pt]{article}

\newfont{\goth}{cmbxti10 scaled\magstep1}
\newcommand{\gotg}{\mbox{\goth g}}
\newcommand{\smin}{\,\raisebox{0.06em}{${\scriptstyle \in}$}\,}
\newcommand{\vef}{\mbox{\LARGE\em -\hspace{-0.42em}x}}

\begin{document}

\renewcommand{\thefootnote}{\fnsymbol{footnote}}

\title{The Algebra of the Energy-Momentum Tensor
       and the Noether Currents in Classical Non-Linear Sigma Models}

\author{M. Forger$\,^{1,2}$ \footnotemark[1]~,~~
        J. Laartz$\,^1$ \footnotemark[2]~,~~
        U. Sch\"aper$\,^1$}
 \footnotetext[1]{Supported by the Deutsche Forschungsgemeinschaft,
                  Contract No.~Ro 864/2-1}
 \footnotetext[2]{Supported by the Studienstiftung des Deutschen Volkes}

\date{\normalsize
      $^1\,$ Fakult\"at f\"ur Physik der Universit\"at Freiburg, \\
      Hermann-Herder-Str.~3, D-7800 Freiburg / FRG \\[0.2cm]
      $^2\,$ Fakult\"at f\"ur Mathematik und Informatik
      der Universit\"at Mannheim, \\
      D-6800 Mannheim / FRG}

\maketitle
\thispagestyle{empty}
\begin{abstract}
\noindent
The recently derived current algebra of classical non-linear sigma models on
arbitrary Riemannian manifolds is extended to include the energy-momentum
tensor. It is found that in two dimensions the energy-momentum tensor
$\theta_{\mu\nu}$, the Noether current $j_\mu$ associated with the global
symmetry of the theory and the composite field $j$ appearing as the
coefficient of the Schwinger term in the current algebra, together with the
derivatives of $j_\mu$ and $j$, generate a closed algebra. The subalgebra
generated by the light-cone components of the energy-momentum tensor consists
of
two commuting copies of the Virasoro algebra, with central charge $\, c\!=\!0
$,
reflecting the classical conformal invariance of the theory, but the current
algebra part and the semidirect product structure are quite different from
the usual Kac-Moody / Sugawara type construction.
\end{abstract}

\vfill

 \begin{flushright}
 \parbox{12em}
 { \begin{center}
 University of Freiburg \\
 THEP 92/24 \\
 October 1992
 \end{center} }
 \end{flushright}

\newpage
\renewcommand{\thefootnote}{\arabic{footnote}}
\setcounter{page}{1}

\noindent In a recent paper \cite{FLS1}, we have derived the current algebra
for classical non-linear sigma models defined on Riemannian manifolds.
This algebra is quite simple to write down and yet does not seem to belong
to any of the algebras which are well known in mathematical physics, mainly
because it involves non-standard (in particular, non-central) extensions of
loop algebras \cite{L}.

On the other hand, the classical non-linear sigma model in two dimensions
is conformally invariant, so its energy-momentum tensor must satisfy the
classical version of the standard commutation relations of conformal field
theory, that is, under Poisson brackets its light-cone components must
generate two commuting copies of the Witt algebra (the Virasoro algebra
with vanishing central charge). We shall verify that this is indeed the
case. Moreover, we shall derive the Poisson bracket relations between the
energy-momentum tensor on the one hand and the Noether currents on the other
hand. The resulting total algebra exhibits, in a concrete field-theoretical
model with continuous internal symmetries, the possibility of reconciling
conformal invariance, expressed through a chiral energy-momentum tensor
algebra, with a non-chiral current algebra, at least at the classical level.

Thus consider the classical two-dimensional non-linear sigma model, whose
configuration space is the space of (smooth) maps $\varphi$ from a given
two-dimensional Lorentz manifold $\Sigma$ to a given Riemannian manifold
$M$, with metric $g$, while the corresponding phase space consists of pairs
$(\varphi,\pi)$ with $\varphi$ as before and $\pi$ a (smooth) section of
the pull-back $\, \varphi^\ast (T^\ast M) \,$ of the cotangent bundle of
$M$ to $\Sigma$ via $\varphi$. The action, written in terms of isothermal
local coordinates $x^\mu$ on $\Sigma$ and of arbitrary local coordinates
$u^i$ on $M$, reads
\begin{equation}
 S~=~{\textstyle{1\over 2}} \int d^2\! x \; \eta^{\mu\nu} \, g_{ij}(\varphi) \,
     \partial_\mu \varphi^i \, \partial_\nu\varphi^j~~,           \label{eq:S1}
\end{equation}
where the $\, \eta_{\mu\nu} \,$ are the coefficients of the standard Minkowski
metric. Thus using a dot to denote the time derivative and a prime to denote
the spatial derivative, we have
\begin{equation}
 \pi_i~=~g_{ij}(\varphi) \, \dot{\varphi}^j~~,                    \label{eq:CM}
\end{equation}
and the canonical Poisson brackets are
\begin{eqnarray}
 \{ \varphi^i(x)\, , \varphi^j(y) \} = 0 &,& \{ \pi_i(x)\, , \pi_j(y) \} =
0~~~,
                                                              \nonumber \\[1mm]
 \{ \varphi^i(x)\, , \pi_j(y) \} &=& \delta^i_j \, \delta(x-y)~~~.
                                                                 \label{eq:CCR}
\end{eqnarray}
The energy-momentum tensor $\theta_{\mu\nu}$ of the theory is most conveniently
derived by variation of the Lagrangian with respect to the metric on $\Sigma$.
(For details, see e.g.\ \cite[p.\ 64 ff]{HE} or \cite[p.\ 504 f]{MTW}.)
It reads
\begin{equation}
 \theta_{\mu\nu}~=~g_{ij}(\varphi) \, \partial_\mu \varphi^i \,
                                      \partial_\nu \varphi^j \,
                   - \, {\textstyle{1\over 2}} \,
                        \eta_{\mu\nu} \, \eta^{\kappa\lambda} \,
                        g_{ij}(\varphi) \, \partial_\kappa \varphi^i \,
                                           \partial_\lambda \varphi^j~~,
                                                                \label{eq:EMT1}
\end{equation}
and it is obviously traceless:
\begin{equation}
 \eta^{\mu\nu} \, \theta_{\mu\nu}~=~0~~.                        \label{eq:TEMT}
\end{equation}
We also assume that the theory exhibits a global invariance under some internal
symmetry group, represented by a (connected) Lie group $G$ acting on $M$ by
isometries, and we shall write $\gotg\,$ for the corresponding Lie algebra
and $\, X_M \smin \, \vef\,(M) \,$ for the fundamental Killing vector field
on $M$ associated with a given generator $\, X\smin\gotg\;$ of $G$:
\begin{equation}
 X_M(m)~=~{d\over dt} \, (\exp(tX) \cdot m) \, \Big|_{t=0}~~.   \label{eq:FVF1}
\end{equation}
Then the Noether current $j_\mu$, taking values in the dual $\gotg^*$ of
$\gotg$,
and the scalar field $j$, taking values in the second symmetric tensor power
$S^2(\gotg^*)$, are given by \cite{FLS1}
\begin{equation}
 \langle j_\mu , X \rangle~=~- \, g_{ij}(\varphi) \, \partial_\mu \varphi^i \,
                                  X_M^j(\varphi)~                \label{eq:NC1}
\end{equation}
for $\, X\smin\gotg\,$, and
\begin{equation}
 \langle j , X\otimes Y \rangle~=~g_{ij}(\varphi) \, X_M^i(\varphi) \,
                                 Y_M^j(\varphi)~                 \label{eq:AF1}
\end{equation}
for $\, X,Y\smin\gotg\,$, where $\langle . , . \rangle$ denotes the natural
pairing between a vector space and its dual.
Now the energy-momentum tensor algebra reads (note $\theta_{11} = \theta_{00}$)
\begin{eqnarray}
 \{ \theta_{00}(x)\, , \theta_{00}(y) \}
 &=& (\theta_{01}(x) + \theta_{01}(y)) \, \delta^\prime(x-y)~~,
                                                         \label{eq:EA1} \\[1mm]
 \{ \theta_{00}(x)\, , \theta_{01}(y) \}
 &=& (\theta_{00}(x) + \theta_{00}(y)) \, \delta^\prime(x-y)~~,
                                                         \label{eq:EA2} \\[1mm]
 \{ \theta_{01}(x)\, , \theta_{01}(y) \}
 &=& (\theta_{01}(x) + \theta_{01}(y)) \, \delta^\prime(x-y)~~,
                                                         \label{eq:EA3}
\end{eqnarray}
while the current algebra is \cite{FLS1}
\begin{eqnarray}
 \{ \langle j_0(x)\,, X \rangle \, , \langle j_0(y)\,, Y \rangle \}
 &=& - \; \langle j_0(x)\,, [X,Y] \rangle \, \delta(x-y)~~,
                                                         \label{eq:CA1} \\[1mm]
 \{ \langle j_0(x)\,, X \rangle \, , \langle j_1(y)\,, Y \rangle \}
 &=& - \; \langle j_1(x)\,, [X,Y] \rangle \, \delta (x-y)           \nonumber
\\
 & & + \; \langle j(y)\,, X\otimes Y \rangle \, \delta^\prime(x-y)~~,
                                                         \label{eq:CA2} \\[1mm]
 \{ \langle j_0(x)\,, X \rangle \, , \langle j(y)\,, Y\otimes Z \rangle \}
 &=& - \; \langle j(x)\;,\, [X,Y]\otimes Z + Y\otimes [X,Z] \rangle \,
     \delta(x-y)~~,~~~~                                  \label{eq:CA3}
\end{eqnarray}
(the remaining Poisson brackets vanish), and the mixed Poisson brackets are
\begin{eqnarray}
 \{ \theta_{00}(x)\, , \langle j_0(y)\,, X \rangle \}
 &=& \langle j_1(x)\,, X \rangle \, \delta^\prime(x-y)~~,\label{eq:EC1} \\[1mm]
 \{ \theta_{00}(x)\, , \langle j_1(y)\,, X \rangle \}
 &=& \langle j_0(x)\,, X \rangle \, \delta^\prime(x-y)             \nonumber \\
 & & - \; \langle ( \partial_0 j_1 - \partial_1 j_0 )(x)\,, X \rangle \,
     \delta(x-y)~~,                                      \label{eq:EC2} \\[1mm]
 \{ \theta_{01}(x)\, , \langle j_0(y)\,, X \rangle \}
 &=& \langle j_0(x)\,, X \rangle \, \delta^\prime(x-y)~~,\label{eq:EC3} \\[1mm]
 \{ \theta_{01}(x)\, , \langle j_1(y)\,, X \rangle \}
 &=& \langle j_1(x)\,, X \rangle \, \delta^\prime(x-y)~~,\label{eq:EC4} \\[1mm]
 \{ \theta_{00}(x)\, , \langle j(y)\,, X\otimes Y \rangle \}
 &=& - \; \langle \partial_0 j(x)\,, X\otimes Y \rangle \,
     \delta(x-y)~~,                                      \label{eq:EC5} \\[1mm]
 \{ \theta_{01}(x)\, , \langle j(y)\,, X\otimes Y \rangle \}
 &=& - \; \langle \partial_1 j(x)\,, X\otimes Y \rangle \,
     \delta(x-y)~~.                                      \label{eq:EC6}
\end{eqnarray}

In higher dimensions, the current algebra (eqns (\ref{eq:CA1}-\ref{eq:CA3}))
and the mixed Poisson brackets (eqns (\ref{eq:EC1}-\ref{eq:EC6})) remain
essentially unchanged, while the energy-momentum tensor algebra (eqns
(\ref{eq:EA1}-\ref{eq:EA3})) is substantially modified and in fact
no longer closes.

In order to prove eqns (\ref{eq:EA1}-\ref{eq:EC6}) in $\, d\!=\!2 \,$ and at
the same time explain what goes wrong for $\, d\!>\!2$, consider the classical
non-linear sigma model over a $d$-dimensional Lorentz manifold $\Sigma$, with
metric $h$, where the formulae for the action and for the energy-momentum
tensor, eqns (\ref{eq:S1}) and (\ref{eq:EMT1}), are replaced by
\begin{equation}
 S~=~{\textstyle{1\over 2}} \int d^d\! x \, \sqrt{\vert {\rm det}(h) \vert} \;
     h^{\mu\nu} \, g_{ij}(\varphi) \,
     \partial_\mu \varphi^i \, \partial_\nu\varphi^j~~,           \label{eq:S2}
\end{equation}
and by
\begin{equation}
 \theta_{\mu\nu}~=~g_{ij}(\varphi) \, \partial_\mu \varphi^i \,
                                      \partial_\nu \varphi^j \,
                   - \, {\textstyle{1\over 2}} \,
                        h_{\mu\nu} \, h^{\kappa\lambda} \,
                        g_{ij}(\varphi) \, \partial_\kappa \varphi^i \,
                                           \partial_\lambda \varphi^j~~,
                                                                \label{eq:EMT2}
\end{equation}
respectively, while most of the other relations given above, namely eqns
(\ref{eq:CM},\ref{eq:CCR}) and (\ref{eq:FVF1}-\ref{eq:AF1}), remain as
they stand. For simplicity, we shall carry out our calculations for the case
where $\Sigma$ is $d$-dimensional Minkowski space, that is, $\; h = \eta =
{\rm diag}(1,-1,...,-1) \,$, which is obviously sufficient when $\, d\!=\!2$,
due to the existence of isothermal local coordinates. The generalization
to arbitrary $d$-dimensional Lorentz manifolds does not present any new
features. To further simplify the calculation, we introduce an auxiliary
field $\tilde{\theta}_{\mu\nu}$ according to
\begin{equation}
 \tilde{\theta}_{\mu\nu}~=~g_{ij}(\varphi) \, \partial_\mu \varphi^i \,
                                              \partial_\nu \varphi^j~~,
                                                              \label{eq:EMTIL1}
\end{equation}
so that
\begin{equation}
 \theta_{\mu\nu}~=~\tilde{\theta}_{\mu\nu} - {\textstyle{1\over 2}} \,
                   \eta_{\mu\nu} \, \tilde{\theta}~~,      \label{eq:EMTEMTIL1}
\end{equation}
where $\tilde{\theta}$ is the trace of $\tilde{\theta}_{\mu\nu}$:
\begin{equation}
 \tilde{\theta}~=~\eta^{\kappa\lambda} \, \tilde{\theta}_{\kappa\lambda}~~.
                                                              \label{eq:TEMTIL}
\end{equation}
In components
\begin{footnote}{We use letters from the beginning of the greek alphabet
to denote spatial indices, running from 1 to $\, d\!-\!1$. In this case no
distinction is made between upper and lower indices, and the usual summation
convention for Euclidean space remains in force.}
\end{footnote},
\begin{eqnarray}
 \theta_{00} &=& {\textstyle{1\over 2}} \, ( \tilde{\theta}_{00} \, + \,
                 \tilde{\theta}_{\gamma\gamma} )~~,     \label{eq:EMTEMTIL2} \\
 \theta_{0\alpha} &=& \tilde{\theta}_{0\alpha}~~,       \label{eq:EMTEMTIL3} \\
 \theta_{\alpha\beta} &=& \tilde{\theta}_{\alpha\beta} \, + \,
                          {\textstyle{1\over 2}} \, \delta_{\alpha\beta} \,
                          ( \tilde{\theta}_{00} \, - \,
                          \tilde{\theta}_{\gamma\gamma} )~~,
                                                        \label{eq:EMTEMTIL4}
\vspace{-1mm}
\end{eqnarray}
or explicitly
\begin{eqnarray}
 \theta_{00} &=& {\textstyle{1\over 2}} \left(
                  g^{ij}(\varphi) \, \pi_i \, \pi_j \, + \,
                  g_{ij}(\varphi) \, \partial_\gamma \varphi^i \,
                                  \partial_\gamma \varphi^j \right)~~,
                                                        \label{eq:EMT3} \\[1mm]
 \theta_{0\alpha} &=& \pi_i \, \partial_\alpha \varphi^i~~,
                                                        \label{eq:EMT4} \\[1mm]
 \theta_{\alpha\beta} &=& g_{ij}(\varphi) \, \partial_\alpha \varphi^i \,
                                             \partial_\beta \varphi^j \, + \,
                          {\textstyle{1\over 2}} \, \delta_{\alpha\beta} \left(
                          g^{ij}(\varphi) \, \pi_i \, \pi_j \, - \,
                          g_{ij}(\varphi) \, \partial_\gamma \varphi^i \,
                                          \partial_\gamma \varphi^j \right)~~.
                                                        \label{eq:EMT5}
\end{eqnarray}
In addition, a composite field containing second (covariant) derivatives
of $\varphi$ will appear:
\begin{equation}
 \tilde{\theta}_{\kappa(\mu\nu)}~
 =~g_{ij}(\varphi) \,
  \partial_\kappa \varphi^i \, \nabla_\mu \partial_\nu \varphi^j~
 =~g_{ij}(\varphi) \, \partial_\kappa \varphi^i \left(
   \partial_\mu \partial_\nu \varphi^j \, + \, {\mit \Gamma}^j_{kl}(\varphi) \,
   \partial_\mu \varphi^k \, \partial_\nu \varphi^l \right)~~.  \label{eq:HCF1}
\end{equation}
Note that the $\tilde{\theta}_{\kappa(\mu\nu)}$ with two indices equal can
be expressed as derivatives of the $\tilde{\theta}_{\mu\nu}$:
\begin{eqnarray}
 \tilde{\theta}_{\mu(\mu\nu)} &=& {\textstyle{1\over 2}} \,
                                 \partial_\nu \tilde{\theta}_{\mu\mu} \qquad
                                 \phantom{\partial_\mu \tilde{\theta}_{\mu\nu}
                                          \, - \, \mbox{}}
                                 {\rm (no~summation)}~~,     \label{eq:HCF2} \\
 \tilde{\theta}_{\nu(\mu\mu)} &=& \partial_\mu \tilde{\theta}_{\mu\nu}
                                 \, - \,  {\textstyle{1\over 2}} \,
                                 \partial_\nu \tilde{\theta}_{\mu\mu} \qquad
                                 {\rm (no~summation)}~~.     \label{eq:HCF3}
\end{eqnarray}
Now we are ready to write down the Poisson brackets involving the auxiliary
field $\tilde{\theta}_{\mu\nu}$. The mixed Poisson brackets are
\begin{eqnarray}
 \{ \tilde{\theta}_{00}(x)\, , \langle j_0(y)\,, X \rangle \}
 &=& 0~~,                                                \label{eq:AM1} \\[1mm]
 \{ \tilde{\theta}_{00}(x)\, , \langle j_\gamma(y)\,, X \rangle \}
 &=& 2 \, \langle j_0(x)\,, X \rangle \, \partial_\gamma \delta(x-y)
                                                                   \nonumber \\
 & & - \; 2 \, \langle ( \partial_0 j_\gamma
                       - \partial_\gamma j_0 )(x)\,, X \rangle \,
     \delta(x-y)~~,                                      \label{eq:AM2} \\[1mm]
 \{ \tilde{\theta}_{0\alpha}(x)\, , \langle j_0(y)\,, X \rangle \}
 &=& \langle j_0(x)\,, X \rangle \, \partial_\alpha \delta(x-y)~~,
                                                         \label{eq:AM3} \\[1mm]
 \{ \tilde{\theta}_{0\alpha}(x)\, , \langle j_\gamma(y)\,, X \rangle \}
 &=& \langle j_\alpha(x)\,, X \rangle \, \partial_\gamma \delta(x-y)
                                                                   \nonumber \\
 & & - \; \langle ( \partial_\alpha j_\gamma
                  - \partial_\gamma j_\alpha )(x)\,, X \rangle \,
     \delta(x-y)~~,                                              \label{eq:AM4}
   \\[1mm]
 \{ \tilde{\theta}_{\alpha\beta}(x)\, , \langle j_0(y)\,, X \rangle \}
 &=& \langle j_\alpha(x)\,, X \rangle \, \partial_\beta \delta(x-y) \, + \,
     \langle j_\beta(x)\,, X \rangle \, \partial_\alpha \delta(x-y)~~,~~~~
                                                         \label{eq:AM5} \\[1mm]
 \{ \tilde{\theta}_{\alpha\beta}(x)\, , \langle j_\gamma(y)\,, X \rangle \}
 &=& 0~~,                                                \label{eq:AM6} \\[1mm]
 \{ \tilde{\theta}_{00}(x)\, , \langle j(y)\,, X\otimes Y \rangle \}
 &=& - \; 2 \, \langle \partial_0 j(x)\,, X\otimes Y \rangle \,
     \delta(x-y)~~,                                      \label{eq:AM7} \\[1mm]
 \{ \tilde{\theta}_{0\alpha}(x)\, , \langle j(y)\,, X\otimes Y \rangle \}
 &=& - \; \langle \partial_\alpha j(x)\,, X\otimes Y \rangle \,
     \delta(x-y)~~,                                      \label{eq:AM8} \\[1mm]
 \{ \tilde{\theta}_{\alpha\beta}(x)\, , \langle j(y)\,, X\otimes Y \rangle \}
 &=& 0~~,                                                \label{eq:AM9}
\end{eqnarray}
and the Poisson brackets between the $\tilde{\theta}_{\mu\nu}$ read
\begin{eqnarray}
 \{ \tilde{\theta}_{00}(x)\, , \tilde{\theta}_{00}(y) \} &=& 0~~,
                                                         \label{eq:AA1} \\[1mm]
 \{ \tilde{\theta}_{00}(x)\, , \tilde{\theta}_{0\alpha}(y) \}
 &=& (\tilde{\theta}_{00}(x) + \tilde{\theta}_{00}(y)) \,
     \partial_\alpha \delta(x-y)~~,                      \label{eq:AA2} \\[1mm]
 \{ \tilde{\theta}_{00}(x)\, , \tilde{\theta}_{\alpha\beta}(y) \}
 &=& 2 \, \tilde{\theta}_{0\alpha}(x) \, \partial_\beta \delta(x-y) \, + \,
     2 \, \tilde{\theta}_{0\beta}(x) \, \partial_\alpha \delta(x-y) \nonumber\\
 & & \makebox[1.2cm]{} + \, 4 \, \tilde{\theta}_{0(\alpha\beta)}(x) \,
     \delta(x-y)~~,                                      \label{eq:AA3} \\[1mm]
 \{ \tilde{\theta}_{0\alpha}(x)\, , \tilde{\theta}_{0\beta}(y) \}
 &=& \tilde{\theta}_{0\beta}(x) \, \partial_\alpha \delta(x-y) \, + \,
     \tilde{\theta}_{0\alpha}(y) \, \partial_\beta \delta(x-y)~~,
                                                         \label{eq:AA4} \\[1mm]
 \{ \tilde{\theta}_{0\gamma}(x)\, , \tilde{\theta}_{\alpha\beta}(y) \}
 &=& \tilde{\theta}_{\gamma\alpha}(x) \, \partial_\beta \delta(x-y) \, + \,
     \tilde{\theta}_{\gamma\beta}(x) \, \partial_\alpha \delta(x-y) \nonumber\\
 & & \makebox[1.2cm]{} + \, 2 \, \tilde{\theta}_{\gamma(\alpha\beta)}(x) \,
     \delta(x-y)~~,                                      \label{eq:AA5} \\[1mm]
 \{ \tilde{\theta}_{\alpha\beta}(x)\, , \tilde{\theta}_{\gamma\delta}(y) \}
 &=& 0~~.                                                \label{eq:AA6}
\end{eqnarray}
The proof goes by explicit computation, using the formulas
\[
 \langle \partial_\mu j_\nu , X \rangle~
 =~ - \; \partial_\mu \partial_\nu\varphi^i \, (X_M)_i(\varphi) \,
    - \, \partial_\mu \varphi^i \, \partial_\nu \varphi^j \,
         (X_M)_{j\vert i}(\varphi)
\]
and
\[
 (f(x) - f(y)) \, \delta^\prime(x-y)~=~- \, f^\prime(x) \, \delta(x-y)~~,
\]
together with the following relations,
\begin{eqnarray*}
 \partial_k g_{ij} \, Z^k &=&- \, g_{ki} \, \partial_j Z^k \,
                             - \, g_{kj} \, \partial_i Z^k~~,                \\
 \partial_k g^{ij} \, Z^k &=&+ \, g^{ki} \, \partial_k Z^j \,
                             + \, g^{kj} \, \partial_k Z^i~~,
\end{eqnarray*}
valid for any Killing vector field $Z$ on $M$ (we omit the argument $x$ or $y$
as soon as there is a factor $\delta(x-y)$):
\vspace{2mm}
\begin{eqnarray*}
\lefteqn{\{ \tilde{\theta}_{00}(x)\, , \langle j_0(y)\,, X \rangle \}~~
   =~~- \; \{ \, g^{ij}(\varphi(x)) \, \pi_i(x) \, \pi_j(x) \, ,
              \, \pi_k(y) \, X_M^k(\varphi(y)) \, \}}      \hspace{9mm} \\[1mm]
  &=& 2 \, g^{ij} \, \pi_j \, \pi_k \, (\partial_i X_M^k) \,
      \delta(x-y)
  ~-~ (\partial_k g^{ij}) \, \pi_i \, \pi_j \, X_M^k \,
      \delta(x-y)                                                       \\[1mm]
  &=& \left( 2 \, g^{ij} \, \pi_k \, (\partial_i X_M^k) \,
   - \, (g^{ik} \, \partial_k X_M^j + g^{jk} \, \partial_k X_M^i)
        \, \pi_i \right) \pi_j \, \delta(x-y)                           \\[1mm]
  &=& 0~~,                                                              \\[3mm]
\lefteqn{\{ \tilde{\theta}_{00}(x)\, , \langle j_\gamma(y)\,, X \rangle \}~~
   =~~- \; \{ \, g^{ij}(\varphi(x)) \, \pi_i(x) \, \pi_j(x) \, ,
              \, (X_M)_k(\varphi(y)) \, \partial_\gamma \varphi^k(y) \, \}}
                                                           \hspace{9mm} \\[1mm]
  &=& - \; 2 \, g^{ij}(\varphi(x)) \, \pi_j(x) \, (X_M)_i(\varphi(y)) \,
      \partial_\gamma \delta(x-y)                                       \\
  & & + \; 2 \, (\partial_i (X_M)_k) \, g^{ij} \, \pi_j \,
      \partial_\gamma \varphi^k \, \delta(x-y)                          \\[1mm]
  &=& - \; 2 \, \pi_i(x) \, X_M^i(\varphi(x)) \, \partial_\gamma \delta(x-y) \\
  & & + 2 ( \, (\partial_i (X_M)_k) - \; (\partial_k (X_M)_i) )
 \, g^{ij} \, \pi_j \, \partial_\gamma \varphi^k \, \delta(x-y) \\[1mm]
  &=& 2 \, \langle j_0(x)\,, X \rangle \, \partial_\gamma \delta(x-y) \,
      - \, 2 \, ((X_M)_{i\vert j} - (X_M)_{j\vert i}) \,
      \partial_0 \varphi^i \, \partial_\gamma \varphi^j \, \delta(x-y)  \\[1mm]
  &=& 2 \, \langle j_0(x)\,, X \rangle \, \partial_\gamma \delta(x-y) \,
      + \, 2 \, \langle (\partial_\gamma j_0
                       - \partial_0 j_\gamma)(x)\,, X \rangle \,
      \delta(x-y)~~,                                                    \\[3mm]
\lefteqn{\{ \tilde{\theta}_{0\alpha}(x)\, , \langle j_0(y)\,, X \rangle \}~~
   =~~- \; \{ \, \pi_i(x) \, \partial_\alpha \varphi^i(x) \, ,
              \, \pi_k(y) \, X_M^k(\varphi(y)) \, \}}      \hspace{9mm} \\[1mm]
  &=& - \; \pi_i(x) \, X_M^i(\varphi(y)) \, \partial_\alpha \delta(x-y) \,
      + \, (\partial_i X_M^k) \, \partial_\alpha \varphi^i \, \pi_k \,
      \delta(x-y)                                                       \\[1mm]
  &=& - \; \pi_i(x) \, X_M^i(\varphi(x)) \, \partial_\alpha \delta(x-y) \\[1mm]
  &=& \langle j_0(x)\,, X \rangle \, \partial_\alpha \delta(x-y)        \\[3mm]
\lefteqn{\{ \tilde{\theta}_{0\alpha}(x)\, ,
            \langle j_\gamma(y)\,, X \rangle \}~~
   =~~- \; \{ \, \pi_i(x) \, \partial_\alpha \varphi^i(x) \, ,
              \, (X_M)_k(\varphi(y)) \, \partial_\gamma \varphi^k(y) \, \}}
                                                           \hspace{9mm} \\[1mm]
  &=& - \; \partial_\alpha \varphi^i(x) \, (X_M)_i(\varphi(y)) \,
      \partial_\gamma \delta(x-y)                                       \\
  & & + \; (\partial_i (X_M)_k) \, \partial_\alpha \varphi^i \,
      \partial_\gamma \varphi^k \, \delta(x-y)                          \\[1mm]
  &=& \langle j_\alpha(x)\,, X \rangle \, \partial_\gamma \delta(x-y) \,
      - \, ((X_M)_{i\vert j} - (X_M)_{j\vert i}) \,
      \partial_\alpha \varphi^i \, \partial_\gamma \varphi^j \,
      \delta(x-y)                                                       \\[1mm]
  &=& \langle j_\alpha(x)\,, X \rangle \, \partial_\gamma \delta(x-y) \,
      + \, \langle (\partial_\gamma j_\alpha
                  - \partial_\alpha j_\gamma)(x)\,, X \rangle \,
      \delta(x-y)~~,                                                    \\[3mm]
\lefteqn{\{ \tilde{\theta}_{\alpha\beta}(x)\, ,
            \langle j_0(y)\,, X \rangle \}~~
   =~~- \; \{ \, g_{ij}(\varphi(x)) \, \partial_\alpha \varphi^i(x) \,
                                       \partial_\beta \varphi^j(x) \, ,
              \, \pi_k(y) \, X_M^k(\varphi(y)) \, \}}      \hspace{9mm} \\[1mm]
  &=& - \; (\partial_k g_{ij}) \,
      \partial_\alpha \varphi^i \, \partial_\beta \varphi^j \,
      X_M^k \, \delta(x-y)                                     \\
  & & - \; g_{ij}(\varphi(x)) \, \partial_\beta \varphi^j(x) \,
      X_M^i(\varphi(y)) \, \partial_\alpha \delta(x-y)                  \\
  & & - \; g_{ij}(\varphi(x)) \, \partial_\alpha \varphi^i(x) \,
      X_M^j(\varphi(y)) \, \partial_\beta \delta(x-y)                   \\[1cm]
  &=& - \; g_{ij}(\varphi(x)) \, \partial_\beta \varphi^j(x) \,
      X_M^i(\varphi(x)) \, \partial_\alpha \delta(x-y)                  \\
  & & - \; g_{ij}(\varphi(x)) \, \partial_\alpha \varphi^i(x) \,
      X_M^j(\varphi(x)) \, \partial_\beta \delta(x-y)                   \\
  & & - \; \left( \, g_{ij} \, \partial_\alpha \varphi^k \,
                \partial_\beta \varphi^j \, (\partial_k X_M^i) \,
           + \, g_{ij} \, \partial_\alpha \varphi^i \,
                \partial_\beta \varphi^k \, (\partial_k X_M^j) \,)\right.  \\
  & & \left.- \; (\partial_k g_{ij}) \,
      \partial_\alpha \varphi^i \, \partial_\beta \varphi^j \,
      X_M^k \right) \, \delta(x-y)                                     \\[1mm]
  &=& \langle j_\alpha(x)\,, X \rangle \, \partial_\beta \delta(x-y) \, + \,
      \langle j_\beta(x)\,, X \rangle \, \partial_\alpha \delta(x-y)~~, \\[3mm]
\lefteqn{\{ \tilde{\theta}_{\alpha\beta}(x)\, ,
            \langle j_\gamma(y)\,, X \rangle \}~~
   =~~- \; \{ \, g_{ij}(\varphi(x)) \, \partial_\alpha \varphi^i(x) \,
                                       \partial_\beta \varphi^j(x) \, ,
              \, (X_M)_k(\varphi(y)) \, \partial_\gamma \varphi^k(y) \, \}}
                                                           \hspace{9mm} \\[1mm]
  &=& 0~~,                                                              \\[5mm]
\lefteqn{\{ \tilde{\theta}_{00}(x)\, , \langle j(y)\,, X\otimes Y \rangle \}}
                                                           \hspace{9mm} \\[1mm]
  &=& \{ \, g^{ij}(\varphi(x)) \, \pi_i(x) \, \pi_j(x) \, ,
         \, g_{kl}(\varphi(y)) \, X_M^k(\varphi(y)) \, Y_M^l(\varphi(y)) \, \}
                                                                        \\[1mm]
  &=& - \; 2 \, (\partial_i (g_{kl} X_M^k Y_M^l)) \,
      \partial_0 \varphi^i \, \delta(x-y)                               \\[1mm]
  &=& - \; 2 \, \langle \partial_0 j(x)\,, X\otimes Y \rangle \,
      \delta(x-y)~~,                                                    \\[3mm]
\lefteqn{\{ \tilde{\theta}_{0\alpha}(x)\, ,
            \langle j(y)\,, X\otimes Y \rangle \}~~
  =~~\{ \, \pi_i(x) \, \partial_\alpha \varphi^i(x) \, ,
       \, g_{kl}(\varphi(y)) \, X_M^k(\varphi(y)) \, Y_M^l(\varphi(y)) \, \}}
                                                           \hspace{9mm} \\[1mm]
  &=& - \; (\partial_i (g_{kl} X_M^k Y_M^l)) \,
      \partial_\alpha \varphi^i \, \delta(x-y)                          \\[1mm]
  &=& - \; \langle \partial_\alpha j(x)\,, X\otimes Y \rangle \,
      \delta(x-y)~~,                                                    \\[3mm]
\lefteqn{\{ \tilde{\theta}_{\alpha\beta}(x)\, ,
            \langle j(y)\,, X\otimes Y \rangle \}}         \hspace{9mm} \\[1mm]
  &=& \{ \, g_{ij}(\varphi(x)) \, \partial_\alpha \varphi^i(x) \,
                                  \partial_\beta \varphi^j(x) \, ,
         \, g_{kl}(\varphi(y)) \, X_M^k(\varphi(y)) \, Y_M^l(\varphi(y)) \, \}
                                                                        \\[1mm]
  &=& 0~~.
\end{eqnarray*}
\begin{eqnarray*}
\lefteqn{\{ \tilde{\theta}_{00}(x)\, , \tilde{\theta}_{00}(y) \}~~
   =~~\{ \, g^{ij}(\varphi(x)) \, \pi_i(x) \, \pi_j(x) \, ,
         \, g^{kl}(\varphi(y)) \, \pi_k(y) \, \pi_l(y) \, \}}
                                                           \hspace{9mm} \\[1mm]
  &=& - \; 2 \left( g^{ij} \, (\partial_i g^{kl}) \,
                    \pi_j \, \pi_k \, \pi_l \, - \,
                    g^{kl} \, (\partial_k g^{ij}) \,
                    \pi_i \, \pi_j \, \pi_l \right) \delta(x-y)         \\[1mm]
  &=& 0~~,                                                              \\[3mm]
\lefteqn{\{ \tilde{\theta}_{00}(x)\, , \tilde{\theta}_{0\alpha}(y) \}~~
   =~~\{ \, g^{ij}(\varphi(x)) \, \pi_i(x) \, \pi_j(x) \, ,
         \, \pi_k(y) \, \partial_\alpha \varphi^k(y) \, \}}
                                                           \hspace{9mm} \\[1mm]
  &=& 2 \, g^{ij}(\varphi(x)) \, \pi_j(x) \, \pi_i(y) \,
      \partial_\alpha \delta(x-y)                                       \\
  & & + \; (\partial_k g^{ij}) \, \pi_i \, \pi_j \,
      \partial_\alpha \varphi^k \, \delta(x-y)                          \\[1mm]
  &=& + \; g^{ij}(\varphi(x)) \, \pi_j(x) \, \pi_i(x) \,
      \partial_\alpha \delta(x-y) \, + \,
      g^{ij} \, \pi_j \, \partial_\alpha \pi_i \, \delta(x-y)  \\
  & & + \; g^{ij}(\varphi(y)) \, \pi_j(y) \, \pi_i(y) \,
      \partial_\alpha \delta(x-y)                                       \\
  & & - \; (\partial_k g^{ij}) \, \partial_\alpha \varphi^k \,
      \pi_j \, \pi_i \, \delta(x-y) \,
      - \, g^{ij} \, \partial_\alpha \pi_j \, \pi_i \, \delta(x-y)  \\
  & & + \; (\partial_k g^{ij}) \, \pi_i \, \pi_j \,
      \partial_\alpha \varphi^k \, \delta(x-y)                          \\[1mm]
  &=& (\tilde{\theta}_{00}(x) + \tilde{\theta}_{00}(y)) \,
      \partial_\alpha \delta(x-y)~~,                                    \\[1cm]
\lefteqn{\{ \tilde{\theta}_{00}(x)\, , \tilde{\theta}_{\alpha\beta}(y) \}~~
   =~~\{ \, g^{ij}(\varphi(x)) \, \pi_i(x) \, \pi_j(x) \, ,
         \, g_{kl}(\varphi(y)) \, \partial_\alpha \varphi^k(y) \,
                                  \partial_\beta \varphi^l(y) \, \}}
                                                           \hspace{9mm} \\[1mm]
  &=& - \; 2 \, g^{ij} \, (\partial_i g_{kl}) \, \pi_j \,
      \partial_\alpha \varphi^k \, \partial_\beta \varphi^l \, \delta(x-y)   \\
  & & + \; 2 \, g^{ij}(\varphi(x)) \, g_{il}(\varphi(y)) \, \pi_j(x) \,
      \partial_\beta \varphi^l(y) \, \partial_\alpha \delta(x-y)        \\
  & & + \; 2 \, g^{ij}(\varphi(x)) \, g_{ki}(\varphi(y)) \, \pi_j(x) \,
      \partial_\alpha \varphi^k(y) \, \partial_\beta \delta(x-y)        \\[1mm]
  &=& 2 \, \pi_i(x) \, \partial_\alpha \varphi^i(x) \,
      \partial_\beta \delta(x-y) \, + \,
      2 \, \pi_i(x) \, \partial_\beta \varphi^i(x) \,
      \partial_\alpha \delta(x-y)                                       \\
  & & + \; 4 \, \pi_i \, \partial_\alpha \partial_\beta \varphi^i \,
\delta(x-y)
    \,
      + \, 4 \, {\mit\Gamma}^i_{kl} \, \partial_\alpha \varphi^k \,
      \partial_\beta \varphi^l \, \delta(x-y)                           \\[1mm]
  &=& 2 \, \tilde{\theta}_{0\alpha}(x) \, \partial_\beta \delta(x-y) \, + \,
      2 \, \tilde{\theta}_{0\beta}(x) \, \partial_\alpha \delta(x-y)    \\
  & & + \; 4 \, \tilde{\theta}_{0(\alpha\beta)}(x) \, \delta(x-y)~~,
\\[3mm]
\lefteqn{\{ \tilde{\theta}_{0\alpha}(x)\, , \tilde{\theta}_{0\beta}(y) \}~~
   =~~\{ \, \pi_i(x) \, \partial_\alpha \varphi^i(x) \, ,
         \, \pi_k(y) \, \partial_\beta \varphi^k(y) \, \}} \hspace{9mm} \\[1mm]
  &=& \partial_\alpha \varphi^i(x) \, \pi_i(y) \,
      \partial_\beta \delta(x-y) \, + \,
      \pi_i(x) \, \partial_\beta \varphi^i(y) \,
      \partial_\alpha \delta(x-y)                                       \\[1mm]
  &=& + \; \partial_\alpha \varphi^i(y) \, \pi_i(y) \,
      \partial_\beta \delta(x-y) \, - \, \pi_i \,
      \partial_\beta \partial_\alpha \varphi^i \, \delta(x-y)           \\
  & & + \; \pi_i(x) \, \partial_\beta \varphi^i(x) \,
      \partial_\alpha \delta(x-y) \, + \, \pi_i \,
      \partial_\alpha \partial_\beta \varphi^i \, \delta(x-y)           \\[1mm]
  &=& \tilde{\theta}_{0\beta}(x) \, \partial_\alpha \delta(x-y) \, + \,
      \tilde{\theta}_{0\alpha}(y) \, \partial_\beta \delta(x-y)~~,      \\[3mm]
\lefteqn{\{ \tilde{\theta}_{0\gamma}(x)\, , \tilde{\theta}_{\alpha\beta}(y)
\}~~
   =~~\{ \, \pi_i(x) \, \partial_\gamma \varphi^i(x) \, ,
         \, g_{kl}(\varphi(y)) \, \partial_\alpha \varphi^k(y) \,
                                  \partial_\beta \varphi^l(y) \, \}}
                                                           \hspace{9mm} \\[1mm]
  &=& - \; (\partial_i g_{kl}) \, \partial_\gamma \varphi^i \,
      \partial_\alpha \varphi^k \, \partial_\beta \varphi^l \, \delta(x-y)   \\
  & & + \; \partial_\gamma \varphi^i(x) \, g_{il}(\varphi(y)) \,
      \partial_\beta \varphi^l(y) \, \partial_\alpha \delta(x-y)        \\
  & & + \; \partial_\gamma \varphi^i(x) \, g_{ki}(\varphi(y)) \,
      \partial_\alpha \varphi^k(y) \, \partial_\beta \delta(x-y)        \\[1mm]
  &=& + \; \partial_\gamma \varphi^i(x) \, g_{il}(\varphi(x)) \,
      \partial_\beta \varphi^l(x) \, \partial_\alpha \delta(x-y)        \\
  & & + \; \partial_\gamma \varphi^i(x) \, g_{ki}(\varphi(x)) \,
      \partial_\alpha \varphi^k(x) \, \partial_\beta \delta(x-y)        \\
  & & + \; 2 \, g_{ij} \, \partial_\gamma \varphi^i \,
      \nabla_\alpha \partial_\beta \varphi^j \, \delta(x-y)             \\[1mm]
  &=& \tilde{\theta}_{\gamma\alpha}(x) \, \partial_\beta \delta(x-y) \, + \,
      \tilde{\theta}_{\gamma\beta}(x) \, \partial_\alpha \delta(x-y) \,
      + \, 2 \, \tilde{\theta}_{\gamma(\alpha\beta)}(x) \,
      \delta(x-y)~~,                                                    \\[3mm]
\lefteqn{\{ \tilde{\theta}_{\alpha\beta}(x)\, ,
            \tilde{\theta}_{\gamma\delta}(y) \}~~
   =~~\{ \, g_{ij}(\varphi(x)) \, \partial_\alpha \varphi^i(x) \,
                                  \partial_\beta \varphi^j(x) \, ,
         \, g_{kl}(\varphi(y)) \, \partial_\gamma \varphi^k(y) \,
                                  \partial_\delta \varphi^l(y) \, \}}
                                                           \hspace{9mm} \\[1mm]
  &=& 0~~.
\end{eqnarray*}
Using the relations (\ref{eq:EMTEMTIL2}-\ref{eq:EMTEMTIL4}), the mixed Poisson
brackets involving the $\theta_{\mu\nu}$ and the currents can now be easily
derived from those involving the $\tilde{\theta}_{\mu\nu}$.
\vspace{2mm}
\begin{eqnarray}
 \{ \theta_{00}(x)\, , \langle j_0(y)\,, X \rangle \}
 &=& \langle j_\alpha(x)\,, X \rangle \, \partial_\alpha \delta(x-y)~~,
                                                         \label{eq:EC7} \\[1mm]
 \{ \theta_{00}(x)\, , \langle j_\gamma(y)\,, X \rangle \}
 &=& \langle j_0(x)\,, X \rangle \, \partial_\gamma \delta(x-y)    \nonumber \\
 & & - \; \langle ( \partial_0 j_\gamma
                  - \partial_\gamma j_0 )(x)\,, X \rangle \,
     \delta(x-y)~~,                                      \label{eq:EC8} \\[1mm]
 \{ \theta_{0\alpha}(x)\, , \langle j_0(y)\,, X \rangle \}
 &=& \langle j_0(x)\,, X \rangle \, \partial_\alpha \delta(x-y)~~,
                                                         \label{eq:EC9} \\[1mm]
 \{ \theta_{0\alpha}(x)\, , \langle j_\gamma(y)\,, X \rangle \}
 &=& \langle j_\alpha(x)\,, X \rangle \, \partial_\gamma \delta(x-y)
                                                                   \nonumber \\
 & & - \; \langle ( \partial_\alpha j_\gamma
                  - \partial_\gamma j_\alpha )(x)\,, X \rangle \,
     \delta(x-y)~~,                                     \label{eq:EC10} \\[1mm]
 \{ \theta_{\alpha\beta}(x)\, , \langle j_0(y)\,, X \rangle \}
 &=& \langle j_\alpha(x)\,, X \rangle \, \partial_\beta \delta(x-y) \, + \,
     \langle j_\beta(x)\,, X \rangle \, \partial_\alpha \delta(x-y)
                                                                   \nonumber \\
 & & - \; \delta_{\alpha\beta} \,
     \langle j_\gamma(x)\,, X \rangle \, \partial_\gamma \delta(x-y)~~,
                                                        \label{eq:EC11} \\[1mm]
 \{ \theta_{\alpha\beta}(x)\, , \langle j_\gamma(y)\,, X \rangle \}
 &=& \delta_{\alpha\beta} \,
     \langle j_0(x)\,, X \rangle \, \partial_\gamma \delta(x-y)    \nonumber \\
 & & - \; \delta_{\alpha\beta}
          \langle ( \partial_0 j_\gamma
                  - \partial_\gamma j_0 )(x)\,, X \rangle \,
     \delta(x-y)~~,                                     \label{eq:EC12} \\[1mm]
 \{ \theta_{00}(x)\, , \langle j(y)\,, X\otimes Y \rangle \}
 &=& - \; \langle \partial_0 j(x)\,, X\otimes Y \rangle \,
     \delta(x-y)~~,                                     \label{eq:EC13} \\[1mm]
 \{ \theta_{0\alpha}(x)\, , \langle j(y)\,, X\otimes Y \rangle \}
 &=& - \; \langle \partial_\alpha j(x)\,, X\otimes Y \rangle \,
     \delta(x-y)~~,                                     \label{eq:EC14} \\[1mm]
 \{ \theta_{\alpha\beta}(x)\, , \langle j(y)\,, X\otimes Y \rangle \}
 &=& - \; \delta_{\alpha\beta} \,
          \langle \partial_0 j(x)\,, X\otimes Y \rangle \,
     \delta(x-y)~~.                                     \label{eq:EC15}
\end{eqnarray}
For the pure energy momentum tensor algebra we have in particular
\begin{eqnarray*}
\{ \theta_{00}(x)\, , \theta_{00}(y) \}~~
  & = & {\textstyle{1\over 4}} \,
      \{ \tilde{\theta}_{00}(x)\, , \tilde{\theta}_{\gamma\gamma}(y) \} \, + \,
      {\textstyle{1\over 4}} \,
      \{ \tilde{\theta}_{\gamma\gamma}(x)\, , \tilde{\theta}_{00}(y) \}
                                                           \hspace{9mm} \\[1mm]
  &=& + \; \tilde{\theta}_{0\gamma}(x) \, \partial_\gamma \delta(x-y) \,
      + \; \tilde{\theta}_{0(\gamma\gamma)}(x) \, \delta(x-y)
\\
  & & - \; \tilde{\theta}_{0\gamma}(y) \, \partial_\gamma \delta(y-x) \,
      - \; \tilde{\theta}_{0(\gamma\gamma)}(y) \, \delta(y-x)
\\[1mm]
  &=& (\theta_{0\gamma}(x) + \theta_{0\gamma}(y)) \,
      \partial_\gamma \delta(x-y)~~,                                    \\[3mm]
\{ \theta_{00}(x)\, , \theta_{0\alpha}(y) \}~~
  & = & {\textstyle{1\over 2}}
      \{ \tilde{\theta}_{00}(x)\, , \tilde{\theta}_{0\alpha}(y) \} \, + \,
      {\textstyle{1\over 2}}
      \{ \tilde{\theta}_{\gamma\gamma}(x)\, , \tilde{\theta}_{0\alpha}(y) \}
                                                           \hspace{9mm} \\[1mm]
  &=& {\textstyle{1\over 2}} \,
      (\tilde{\theta}_{00}(x) + \tilde{\theta}_{00}(y)) \,
      \partial_\alpha \delta(x-y)                                            \\
  & & - \; \tilde{\theta}_{\alpha\gamma}(y) \, \partial_\gamma \delta(y-x) \,
      - \; \tilde{\theta}_{\alpha,\gamma\gamma}(y) \, \delta(y-x)       \\[1mm]
  &=& {\textstyle{1\over 2}} \,
      (\tilde{\theta}_{00}(x) + \tilde{\theta}_{00}(y)) \,
      \partial_\alpha \delta(x-y)                                            \\
  & & + \; \tilde{\theta}_{\alpha\gamma}(x) \, \partial_\gamma \delta(x-y) \,
      + \; {\textstyle{1\over 2}} \,
      \partial_\alpha \tilde{\theta}_{\gamma\gamma}(x) \, \delta(x-y)   \\[1mm]
  &=& {\textstyle{1\over 2}} \, ( \,
      \tilde{\theta}_{00}(x) + \tilde{\theta}_{00}(y) -
      \tilde{\theta}_{\gamma\gamma}(x) + \tilde{\theta}_{\gamma\gamma}(y)
      \, ) \, \partial_\alpha \delta(x-y)                                    \\
  & & + \; \tilde{\theta}_{\alpha\gamma}(x) \, \partial_\gamma \delta(x-y)
                                                                        \\[1mm]
  &=& \theta_{00}(y) \, \partial_\alpha \delta(x-y) \, + \,
      \theta_{\alpha\gamma}(x) \, \partial_\gamma \delta(x-y)~~,
\end{eqnarray*}
so
\begin{eqnarray}
 \{ \theta_{00}(x)\, , \theta_{00}(y) \}
 &=& (\theta_{0\gamma}(x) + \theta_{0\gamma}(y)) \,
     \partial_\gamma \delta(x-y)~~,                      \label{eq:EA4} \\[1mm]
 \{ \theta_{00}(x)\, , \theta_{0\alpha}(y) \}
 &=& \theta_{00}(y) \, \partial_\alpha \delta(x-y) \, + \,
     \theta_{\alpha\gamma}(x) \, \partial_\gamma \delta(x-y)~~,
                                                         \label{eq:EA5} \\[1mm]
 \{ \theta_{0\alpha}(x)\, , \theta_{0\beta}(y) \}
 &=& \theta_{0\beta}(x) \, \partial_\alpha \delta(x-y) \, + \,
     \theta_{0\alpha}(y) \, \partial_\beta \delta(x-y)~~,
                                                         \label{eq:EA6}
\end{eqnarray}
plus three other Poisson bracket relations which contain the composite field
$\tilde{\theta}_{\kappa(\mu\nu)}$: these are rather complicated and not very
enlightening, so we shall not write them down explicitly. When $\, d\!=\!2$,
eqns (\ref{eq:EC7}-\ref{eq:EC15}) and (\ref{eq:EA4}-\ref{eq:EA6}) clearly
reduce to eqns (\ref{eq:EC1}-\ref{eq:EC6}) and (\ref{eq:EA1}-\ref{eq:EA3}),
respectively; moreover, at least two of the three indices of the composite
field $\tilde{\theta}_{\kappa(\mu\nu)}$ are necessarily equal, so that,
according to eqns (\ref{eq:HCF2}) and (\ref{eq:HCF3}), it can be
completely eliminated in favor of derivatives of the energy-momentum
tensor, and the resulting formulas turn out to carry no information beyond
eqns (\ref{eq:EA4}-\ref{eq:EA6}) given above. When $ \, d\!>\!2$, however,
there is no way to eliminate this composite field, so the algebra does not
close. Moreover, repeating the process of computing Poisson brackets will
give rise to new composite fields built from products of higher (covariant)
derivatives of $\varphi$. We suspect that a finite number of such additional
fields will suffice to close the algebra (when derivatives are included),
but we have not analyzed the question any further.

Returning to the two-dimensional case, it remains to be shown that eqns
(\ref{eq:EA1}-\ref{eq:EA3}) give rise to a chiral Witt algebra. To do so,
we first switch to light-cone components,
\begin{eqnarray}
 \theta_{++} &=& {\textstyle{1\over 2}} \, (\theta_{00} + \theta_{01})~~,    \\
 \theta_{--} &=& {\textstyle{1\over 2}} \, (\theta_{00} - \theta_{01})~~,
\end{eqnarray}
(note that $\; \theta_{+-} = 0 \,$), and then convert the equal-time Poisson
bracket relations (\ref{eq:EA1}-\ref{eq:EA3}) to commutation relations valid
on the light cone: this requires invoking the equations of motion for the
energy momentum tensor. These are nothing but the conservation law
\begin{equation}
 \partial_-\theta_{++}~=~0~~~,~~~\partial_+\theta_{--}~=~0~~~,  \label{eq:EMTC}
\end{equation}
so $\theta_{++}$ depends only on $x^+$ and $\theta_{--}$ depends only on $x^-$,
then eqns (\ref{eq:EA1}-\ref{eq:EA3}) become
\begin{eqnarray}
 \{ \theta_{++}(x^+)\, , \theta_{++}(y^+) \}
 &=& + \; (\theta_{++}(x^+) + \theta_{++}(y^+)) \,
     \delta^\prime(x^+ - y^+)~~,                                            \\
 \{ \theta_{--}(x^-)\, , \theta_{--}(y^-) \}
 &=& - \; (\theta_{--}(x^-) + \theta_{--}(y^-)) \,
     \delta^\prime(x^- - y^-)~~,                                            \\
 \{ \theta_{++}(x^+)\, , \theta_{--}(y^-) \} &=& 0~~.
\end{eqnarray}
When expressed in Fourier components this becomes the well-known Witt algebra
for $\theta_{++}$ and for $\theta_{--}$. \\
For the currents, on the other hand, this procedure cannot be carried out
in the same way, because the conservation law for the currents is, in
light-cone components
$$
 \partial_+j_- + \partial_-j_+ = 0~~,
$$
and this alone is not sufficient to convert equal-time Poisson brackets to
light-cone Poisson brackets.

The net result is that the total algebra is a semidirect product of a
chiral Witt algebra with a non-chiral (non Kac-Moody) current algebra.

The most important question to be answered next is how the algebraic structure
derived above changes when passing from the classical theory to the quantum
theory. Some changes are to be expected due to the phenomenon of dynamical
mass generation, which will give the energy-momentum tensor a non-vanishing
trace and destroy the chiral nature of the energy-momentum tensor algebra.
Our hope is that the non-chiral current algebra should give a clue as to what
should be this non-chiral energy-momentum tensor algebra, and ideally that
there should exist some non-chiral analogue of the Sugawara construction --
which, as is well known, e.g., for the Wess-Zumino-Novikov-Witten models or
for certain massless fermionic theories \cite{GO}, actually derives the chiral
energy-momentum tensor algebra of conformal field theory from the corresponding
chiral current algebra (two commuting copies of the relevant (untwisted) affine
Kac-Moody algebra).

\end{document}